\documentclass[doublecol]{epl2}
\usepackage{amsmath,amssymb}
\usepackage{color}
\usepackage{graphicx}
\usepackage[utf8]{inputenc}
\usepackage{subfig}

\newcommand{\Sabs}{\dot{S}^\ab{abs}}

\newcommand{\Sins}{\dot{S}^\ab{ins}}
\newcommand{\Smed}{\dot{S}^\ab{m}}
\newcommand{\Wout}{\dot{W}^\ab{out}}
\newcommand{\Srst}{\dot{S}^\ab{rst}}
\newcommand{\Ssys}{\dot{S}^\ab{sys}}

\newcommand{\dWext}{\dot{W}^\ab{ext}}
\newcommand{\dWout}{\dot{W}^\ab{out}}

\newcommand{\intt}{\int_0^\infty \diff{t}}
\newcommand{\intx}{\int \diff{x}}
\newcommand{\diff}[1]{\mathrm{d}#1\;}

\DeclareMathOperator{\sgn}{sgn}

\begin{document}

\title{Stochastic thermodynamics of resetting}

\author{Jaco Fuchs\thanks{These authors contributed equally to this work.} \and Sebastian Goldt$^{(\mathrm{a})}$ \and Udo Seifert\thanks{\email{useifert@theo2.physik.uni-stuttgart.de}}}
\shortauthor{J. Fuchs \etal}

\institute{                    
  II. Institut für Theoretische Physik, Universität Stuttgart -- 70550 Stuttgart, Germany
}


\abstract{Stochastic dynamics with random resetting leads to a non-equilibrium
  steady state. Here, we consider the thermodynamics of resetting by deriving
  the first and second law for reset processes far from equilibrium. We identify
  the contributions to the entropy production of the system which arise due to
  resetting and show that they correspond to the rate with which information is
  either erased or created. Using Landauer's principle, we derive a bound on the
  amount of work that is required to maintain a resetting process. We discuss
  different regimes of resetting, including a Maxwell's demon scenario where
  heat is extracted from a bath at constant temperature.}

\maketitle

\section{Introduction} Stochastic processes with resetting, that is a sudden
transition to a single preselected state or region in phase space, have
attracted a lot of interest recently. They arise in a range of problems from the
optimisation of search strategies (where resetting events restart the search
and can lead to shorter search times~\cite{Evans2011a, Kusmierz2014,
  Gupta2014,meylahn2015,eule2015a}) to kinetic proofreading
\cite{Hopfield1974,Ninio1975,murugan2014,hartich2015a} or even population
dynamics~\cite{Kyriakidis1994,Dharmaraja2015,eule2015a} (where a reset
corresponds to a catastrophic event), to name but a few.

From a thermodynamic perspective, resetting raises interesting issues for three
reasons. First, resetting changes the information content of the system: once the system
is reset, all knowledge about its previous state is lost. Information, however,
is physical and deleting it comes at a thermodynamic cost, an idea that goes
back to Landauer's principle and Bennett's
work~\cite{Landauer1961,Bennett1982}. This connection between information and
thermodynamics has attracted a lot of attention from
stochastic thermodynamics over the last decade, both
theoretically~\cite{Seifert2012,Parrondo2015} and
experimentally~\cite{Berut2012,Jun2014}. Here, the question is what
thermodynamic cost is at least required to implement resetting in a steady
state. Second, a particularly intriguing application of stochastic
thermodynamics is analysing the efficiency of computation in biological systems
\cite{lan2012,barato2014a,hartich2016}. The key idea here is that the
informational efficiency of, say, a search strategy has to be weighted against
the thermodynamic costs of implementing it. Given the apparent role of resetting
processes in nature, an understanding of the dissipation involved is thus
paramount. Finally, the competition between the stochastic exploration of states
and resetting, which constrains the system to a particular region of its phase
space, gives rise to a non-equilibrium steady state (NESS). Such a state is
characterised by a stationary probability distribution and a non-vanishing
probability current and is an intriguing object of study in its own right.

In this letter, we address these points by analysing both, the thermodynamics of
resetting a single colloidal particle, arguably \emph{the} paradigm for the
field~\cite{Evans2011a,Evans2011,Pal2015}, and resetting a discrete system. We
will derive the first and second law of thermodynamics in both cases and
identify contributions to the well-established total entropy production rate of
stochastic thermodynamics~\cite{Seifert2012} which arise due to resetting. We
illustrate the physical interpretation of these rates and their properties using
several examples, including an implementation of Maxwell's demon that extracts
heat from a heat bath at constant temperature using resetting.

\section{Thermodynamics of resetting}

\begin{figure}
  \centering
  \includegraphics[scale=0.9]{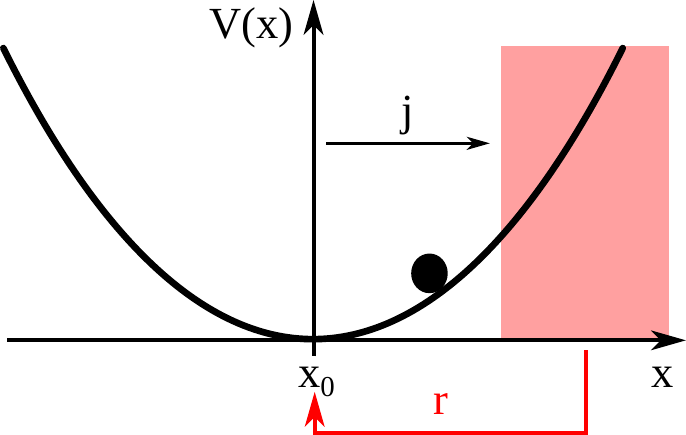}
  \caption{\label{fig:sketch_continuous} Paradigmatic system with
      resetting. A colloidal particle in a potential $V(x)$ at temperature $T$
    is reset to the origin $x_0=0$ at a space-dependent rate $r(x)$, resulting
    in a probability current $j$ from the origin to the region of resetting.}
\end{figure}

We consider an overdamped colloidal particle along a spatial coordinate $x$
immersed in a heat bath at
temperature $T$. The particle experiences a systematic force
$F(x)=-\partial_x V(x)$ and is randomly reset to a fixed position $x_0$ with a
space-dependent rate $r(x)\ge 0$, as shown schematically in
fig. \ref{fig:sketch_continuous}.  The dynamic of the particle is captured by
an augmented Fokker-Planck equation \cite{Gardiner1983},
\begin{multline}
  \label{eq:fpe}
  \partial_t p(x)=-\partial_xj(x) - r(x)p(x) \\
    + \delta(x-x_0)  \int \diff{x'} r(x')p(x') = 0
\end{multline}
where 
\begin{equation}
  j(x) \equiv F(x) p(x)-\partial_x p(x)
\end{equation}
is the probability current. The second and third term on the right-hand side of
eq.~\eqref{eq:fpe} add the probability flux out of each point $x$ and into the
reset position $x_0$ to the standard Fokker-Planck equation, ensuring that
probability is conserved. This interplay of drift and diffusion on the one hand
and resetting on the other leads to a non-equilibrium steady state, where the
probability distribution $p(x)$ of the particle's position is stationary, but
there still is a non-vanishing current $j(x)$. Here and for the remainder of the
letter, we set $T=1$ and choose dimensionless units, without loss of
generality. For general $r(x)$, eq.~\eqref{eq:fpe} has analytical solutions only
in a few cases.

To obtain the first law for the system, we multiply eq.~\eqref{eq:fpe} with
$V(x)$ and integrate by parts, using $\partial_xV(x)=-F(x)$, to find that
\begin{equation}
  \label{eq:first-law_integrals}
  \intx j(x)F(x)+\intx r(x)p(x)\left[V(x)-V(x_0)\right]=0
\end{equation}
Following~\cite{Seifert2012}, we identify the first term in eq.
\eqref{eq:first-law_integrals} as the rate of heat dissipation in the medium
and the second term as the work which is extracted from the
system in the steady state,
\begin{align}
  \dot{Q}     \equiv &  \intx j(x)F(x), \label{eq:heat_cont}\\
  \Wout \equiv &  \intx r(x)p(x)\left[V(x)-V(x_0)\right] \label{eq:Wout_cont}
\end{align}
such that eq.~\eqref{eq:first-law_integrals} can be written as
\begin{equation}
  \label{eq:first-law}
  \dot{Q}+\Wout=0
\end{equation}
which we interpret as the first law of thermodynamics for resetting in the
continuous case.

The (Shannon) entropy of the system is defined as 
\begin{equation}
  \label{eq:Ssys_cont}
  S^\ab{sys}=-\intx p(x)\ln p(x).
\end{equation}
and its derivative $\Ssys=0$ in the steady state.  Differentiating
eq.~\eqref{eq:Ssys_cont} with respect to time and inserting eq.~\eqref{eq:fpe}
yields
\begin{multline}
  \label{eq:second-law_integrals}
    \intx \frac{j(x)^2}{p(x)} =\intx j(x)F(x) -\intx
    r(x)p(x)\ln p(x)  \\
    + \ln p(x_0) \intx r(x)p(x) \ge 0.
\end{multline}
Following~\cite{Seifert2012}, we identify
\begin{equation}
  \label{eq:Smed_cont}
  \Smed \equiv \dot{Q}=\intx j(x)F(x)
\end{equation}
as the entropy production in the surrounding medium and note that it equals the
dissipated heat~\eqref{eq:heat_cont}, as
expected. The last two terms on the right hand side of eq.
\eqref{eq:second-law_integrals} arise from the resetting terms in the
Fokker-Planck equation \eqref{eq:fpe}, suggesting the definition
\begin{equation}
  \label{eq:Sabs_cont}
  \Sabs \equiv \intx r(x)p(x)\ln p(x)
\end{equation}
as the absorption entropy rate, corresponding to the change of Shannon
entropy of the system due to probability flux out of each point $x$. On the
other hand,
\begin{equation}
  \label{eq:Sins_cont}
  \Sins \equiv -\ln p(x_0) \intx r(x)p(x)
\end{equation}
is the insertion entropy rate. It depends on the probability density at the position
$x_0$ to which the particle is reset and the probability flux out of every other
$x$. We call their sum the resetting entropy production rate
\begin{equation}
  \label{eq:Srst_cont}
  \Srst \equiv \Sabs+\Sins=\intx r(x)p(x)\ln \frac{p(x)}{p(x_0)}
\end{equation}
Equality in eq.~\eqref{eq:second-law_integrals} is reached for a vanishing
current only. However, as soon as $r(x)$ is nonzero somewhere, there will be
resetting which directly leads to a nonzero current. Thus
\begin{equation}
  \label{eq:second-law_cont}
  \Smed-\Sabs-\Sins=\Smed-\Srst> 0
\end{equation}
for a non-vanishing reset rate $r(x)$, which we interpret as the second law of
thermodynamics including resetting.

\subsection{Discrete dynamics}

Resetting can also be implemented in discrete systems. We distinguish
two types of transition rates from state $m$ to state $n$. For any connected
states $m$ and $n$, transitions occur at a rate $w_{mn}$, which for
thermodynamic consistency obey
\begin{equation}
  \label{eq:dbc}
  \ln \frac{w_{mn}}{w_{nm}}=E_m-E_n,
\end{equation}
where $E_n$ is the energy of state $n$. Furthermore, resetting from any state
$m$ to a fixed state $n_0$ is performed at a state-dependent reset rate
$r_{m}$. The master equation is then given by
\begin{align}
  \label{eq:master}
    \partial_tp_n=&\sum_m \left[ p_m w_{mn} - p_n w_{nm}\right] \\
                  & \qquad -p_n r_{n} +\delta_{n n_0} \sum_m p_m r_{m} = 0
\end{align}
where the last two terms on the right hand side account for the resetting,
similarly to the augmented Fokker-Planck equation~\eqref{eq:fpe}. However, we
can reduce eq.~\eqref{eq:master} to a standard master equation by introducing
transition rates $w_{mn}'\equiv w_{mn}+r_{m}\delta_{nn_0}$.

Multiplying eq.~\eqref{eq:master} with $E_n$ and summing over all states, we
obtain
\begin{equation}
  \label{eq:first-law_sums}
  \sum_{mn} p_m w_{mn} \ln\frac{w_{mn}}{w_{nm}} +\sum_nr_{n}p_n (E_n-E_{n_0}) =0
\end{equation}
where we have used eq.~\eqref{eq:dbc}. We identify the first term with the
rate of heat dissipation and, hence, the rate of entropy production in the
medium~\cite{Seifert2012}
\begin{equation}
  \label{eq:Smed_disc}
  \dot{Q}  = \Smed \equiv \sum_{mn} p_m w_{mn} \ln \frac{w_{mn}}{w_{nm}}.
\end{equation}
We consequently define the work which is extracted from the system as
\begin{equation}
\label{eq:Wout_disc}
  \dWout \equiv \sum_nr_{n}p_n \left(E_n-E_{n_0}\right)
\end{equation}
such that eq.~\eqref{eq:first-law_sums} becomes the first law eq.~\eqref{eq:first-law}
for resetting processes in the discrete case.

Starting from the discrete Shannon entropy of the system, we
have
\begin{equation}
  \label{eq:Ssys_disc}
  \Ssys=-\sum_n \dot{p}_n \ln p_n = 0.
\end{equation}
Using the master equation~\eqref{eq:master} and similar arguments as before, it
follows that
\begin{multline}
  \label{eq:second-law_sums}
    \sum_{mn}p_m w_{mn}\ln \frac{p_m w_{mn}}{p_n w_{nm}} = \sum_{mn}p_m
    w_{mn}\ln \frac{w_{mn}}{w_{nm}}\\
    \qquad -\sum_n r_{n} p_n\ln p_n+\sum_n r_{n} p_n \ln p_{n_0} \ge 0
\end{multline}
where we have applied the log sum inequality. We define the rate of
entropy production due to absorption and insertion as
\begin{align}
  \Sabs &\equiv \sum_n r_{n} p_n\ln p_n, \label{eq:Sabs_disc}\\
  \Sins &\equiv -\sum_n r_{n} p_n \ln p_{n_0} \label{eq:Sins_disc}
\end{align}
which should be compared to~\eqref{eq:Sabs_cont} and
\eqref{eq:Sins_cont}. Introducing the resetting entropy production for discrete
systems,
\begin{equation}
  \label{eq:Srst_disc}
  \Srst\equiv \sum_n r_{n} p_n\ln \frac{p_n}{p_{n_0}},
\end{equation}
allows us to rewrite eq.~\eqref{eq:second-law_sums} as
\begin{equation}
  \label{eq:second-law_disc}
  \Smed-\Sabs-\Sins\equiv\Smed-\Srst\ge 0.
\end{equation}
This is the second law of thermodynamics for a discrete system with resetting.

We finally note that for time-dependent dynamics $\partial_t p(x,t) \neq 0$
in the continuous and $\partial_t p_n(t) \neq 0$  in the discrete case, respectively, leading to $\Ssys \neq 0$ in
contrast to eq.~\eqref{eq:Ssys_disc}. However, the derivation of the second law
can be written analogously and it hence reads
\begin{equation}
\label{eq:second-law_time}
  \Ssys+\Smed-\Srst \ge 0.
\end{equation}
for both continuous and discrete dynamics.

\subsection{Resetting entropy rate and Landauer's principle}

Resetting can decrease or increase the Shannon entropy compared to steady
states without resetting, if they exist, depending on whether the distribution
$p(x)$ is compressed or broadened.  For example, resetting a freely diffusing
particle to some point $x_0$ concentrates the particle to that region and hence
reduces the uncertainty about its position, resulting in a resetting entropy
rate $\Srst<0$. Specifically, for discrete dynamics we have $p_n\le1$ and hence
$\Sabs<0$, which reduces the system entropy due to the flux
of probability out of each state $n$. For the continuous case, no general
statement about $\Sabs$ can be made. On the other hand, for discrete dynamics,
$\Sins>0$, while it can be both positive or negative for continuous dynamics,
depending on the stochastic entropy of the designated reset state,
$-\ln p(x_0)$.  So inserting into a state with \emph{low} steady state
probability increases the system entropy, $\Srst>0$, while inserting into a
state with \emph{high} probability effectively erases information from the
system, $\Srst<0$.

Let us expand on this last point by analysing the following toy model for
erasure. We consider a two-state system with equal energy levels and a reset
rate $r$ from state 2 to state 1, but no thermal transition rates
$w_{mn}$. Starting from the equilibrium state with $p_1=p_2=1/2$ and Shannon
entropy $S^\ab{sys}=\ln 2$, the probability $p_2$ decreases exponentially,
flowing to state $1$ and leaving the system with $S^\ab{sys}=0$, implementing
the erasure of a single bit. There is no thermodynamic entropy production since
$w_{mn}=0$, but the rate of change of the system's entropy $\Ssys\neq0$ and the
second law \eqref{eq:second-law_time} reduces to $\Ssys=\Srst$, with equality
due to the vanishing rates $w_{mn}$,
\emph{cf}.~\eqref{eq:second-law_sums}. Integrating leads to
\begin{equation}
  \intt \Srst  = \Delta S^\ab{rst}=\Delta S^\ab{sys}=-\ln 2
\end{equation}
which is exactly minus the minimal amount of work needed to erase one bit originally
derived by Landauer~\cite{Landauer1961}. A natural question is now how the
resetting entropy production is related to Landauer's principle more generally.

On the single trajectory level, stochastic entropy is defined as
$s(t)\equiv -\ln p_{n(t)}$~\cite{seifert2005a} in the discrete case and analogously for
continuous dynamics. Each time a reset from state $n$ takes place, the system's
entropy changes by $\Delta s_n=-\ln p_{n_0} +\ln p_n$ and the energy of the system
changes by $\Delta E_n = E_{n_0}-E_n$, corresponding to minus the extracted work.
The difference in non-equilibrium free energy then reads
\begin{equation}
  \Delta F_n = \Delta E_n-\Delta s_n  =E_{n_0}-E_n-\ln \frac{p_n}{p_{n_0}}.
\end{equation}
Moving on to the ensemble level, we average using $r_n p_n$ and use the second
law of non-equilibrium thermodynamics~\cite{Esposito2011,Parrondo2015} to find
the following lower bound for the amount of \emph{external} work necessary to
maintain the resetting:
\begin{equation}
  \begin{split}
  \dWext &\ge \sum_n r_n p_n \Delta F_n \\
  &= \sum_n r_n p_n (E_{n_0}-E_n)-\sum_n r_n p_n \ln \frac{p_n}{p_{n_0}}.
  \end{split}
\end{equation}
We can now identify the extracted work $\Wout$ \eqref{eq:Wout_disc} and
resetting entropy production \eqref{eq:Srst_disc}, such that
\begin{equation}
  \dWext \ge -\Wout -\Srst. \label{eq:landauer}
\end{equation}
We note that $\dWext$ needs to be provided by the resetter, but is not performed
on the system and thus has no influence on the first law \eqref{eq:first-law}.
This result gives a lower bound for the work $\dWext$ which must be performed by
an external mechanism to maintain the resetting process.  Comparing
\eqref{eq:landauer} to the non-equilibrium Landauer principle
\cite{Esposito2011,Parrondo2015}, we thus find that the resetting entropy
production equals the rate with which information is erased from or created in
the system. Hence, \eqref{eq:landauer} implements Landauer's principle in a
steady state, providing a scheme complementary to autonomous demons involving a
tape as information reservoir \cite{mandal2012,barato2013b,barato2014}.
Two regimes emerge from this result: to
continuously reduce the information content of the system in the steady state, we must apply work .
However, we can also extract work by increasing the system entropy, leading to a
Maxwell demon. 

\section{Examples}

We illustrate crucial properties of the entropy rates that we have introduced with simple examples.
First we revisit the freely diffusing particle with resetting and
find that $\Srst\le0$, as we do for diffusion with drift.
However, this inequality is not a general result as we
show in our third example. We also demonstrate that we can extract heat from the
bath using a reset mechanism and give a full discussion of the different regimes
that emerge in our final example.

\begin{figure}
  \centering
   \subfloat[Free diffusion with reset rate
  $r=6$ (scaled with $1/30$) for $|x|>1$]{
      \label{fig:example_window}
      \includegraphics[width=.33\textwidth]{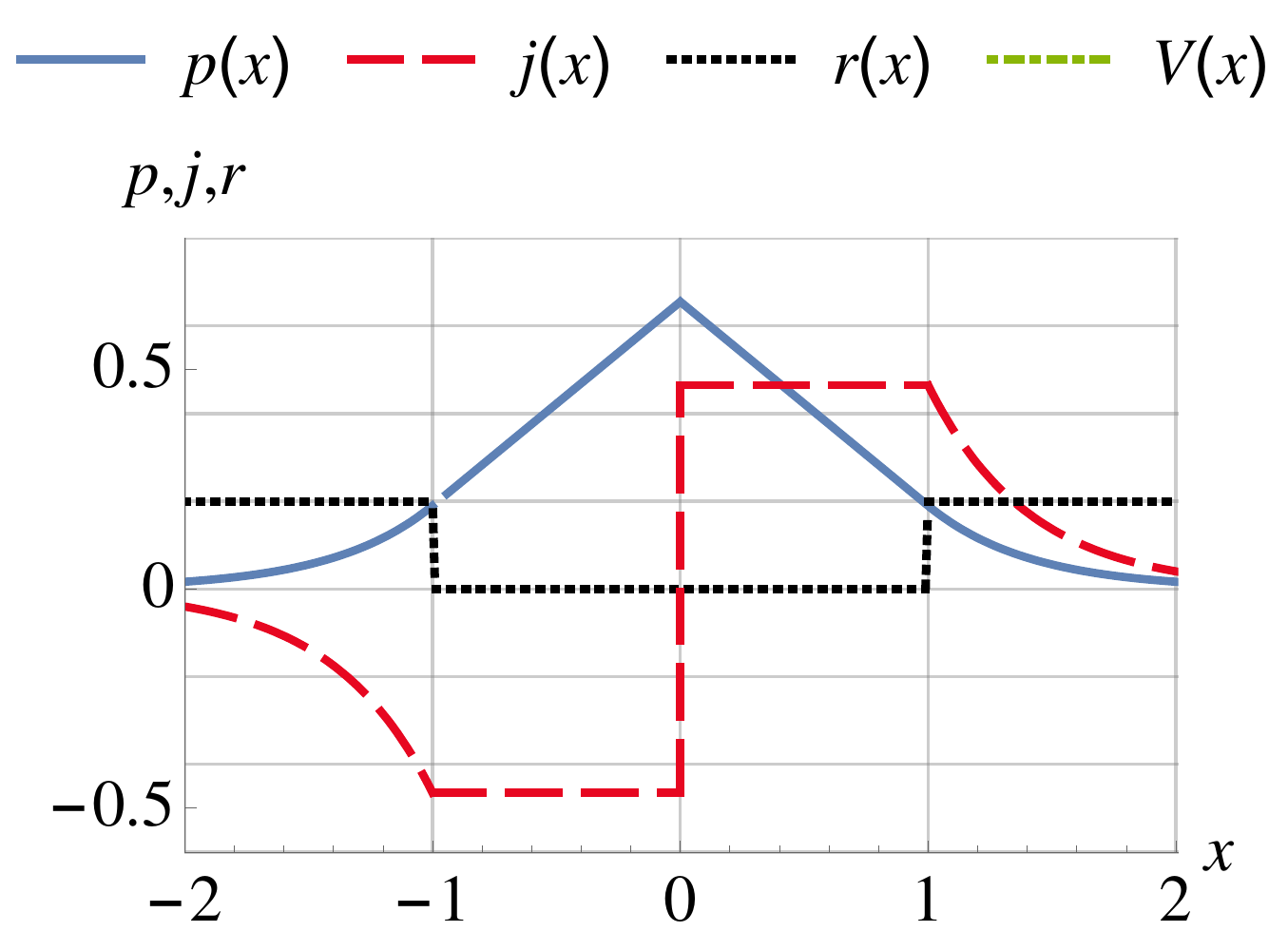}
    }\quad
    \subfloat[Diffusion in a V-shaped potential $V(x)$ (scaled with
    $1/3$) with constant reset rate $r=1$ (scaled with $1/10$).]{
      \label{fig:example_v}
      \includegraphics[width=.33\textwidth]{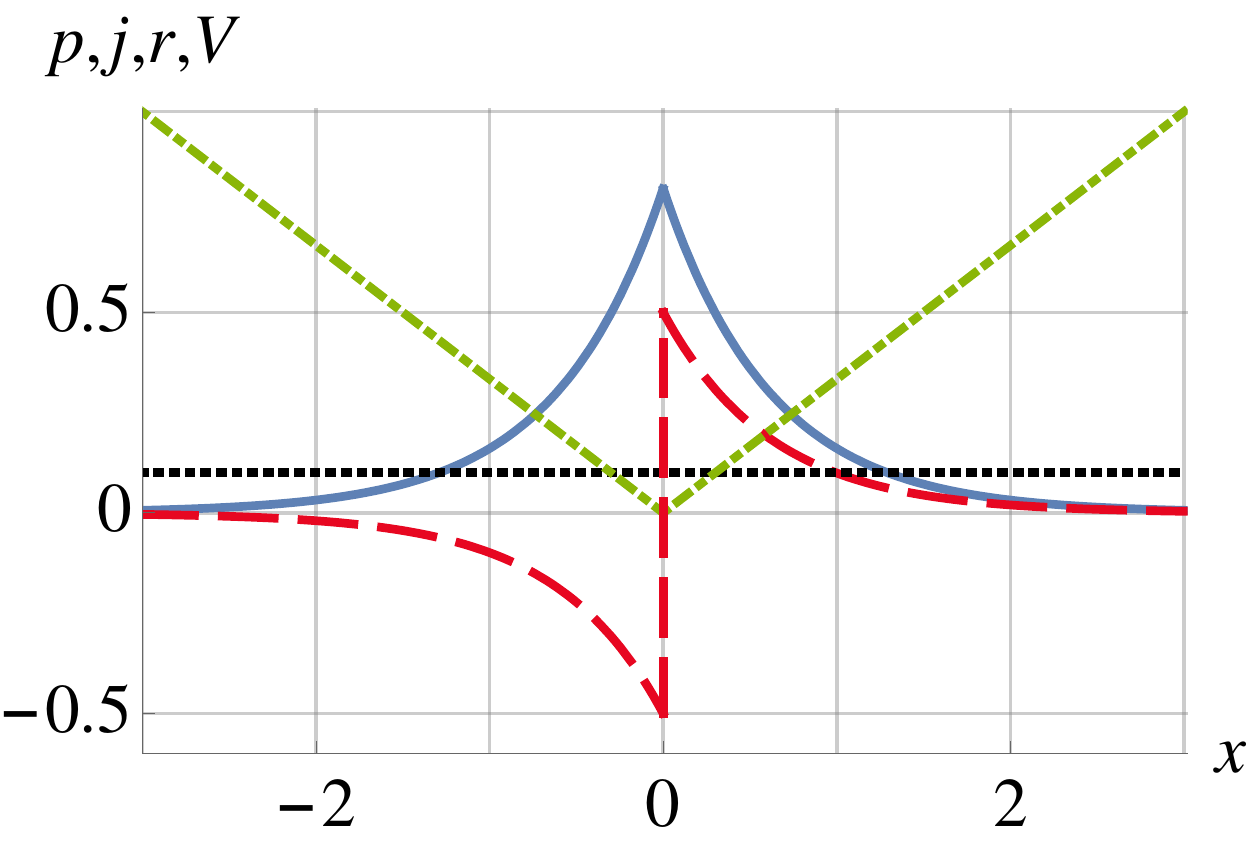}
    }\quad
    \subfloat[Diffusion in a potential well with a reflecting boundary 
    at $x=1$ and resetting at constant rate $r=1$ (scaled with $1/5$)
    for $0<x<1$.]{
      \includegraphics[width=.33\textwidth]{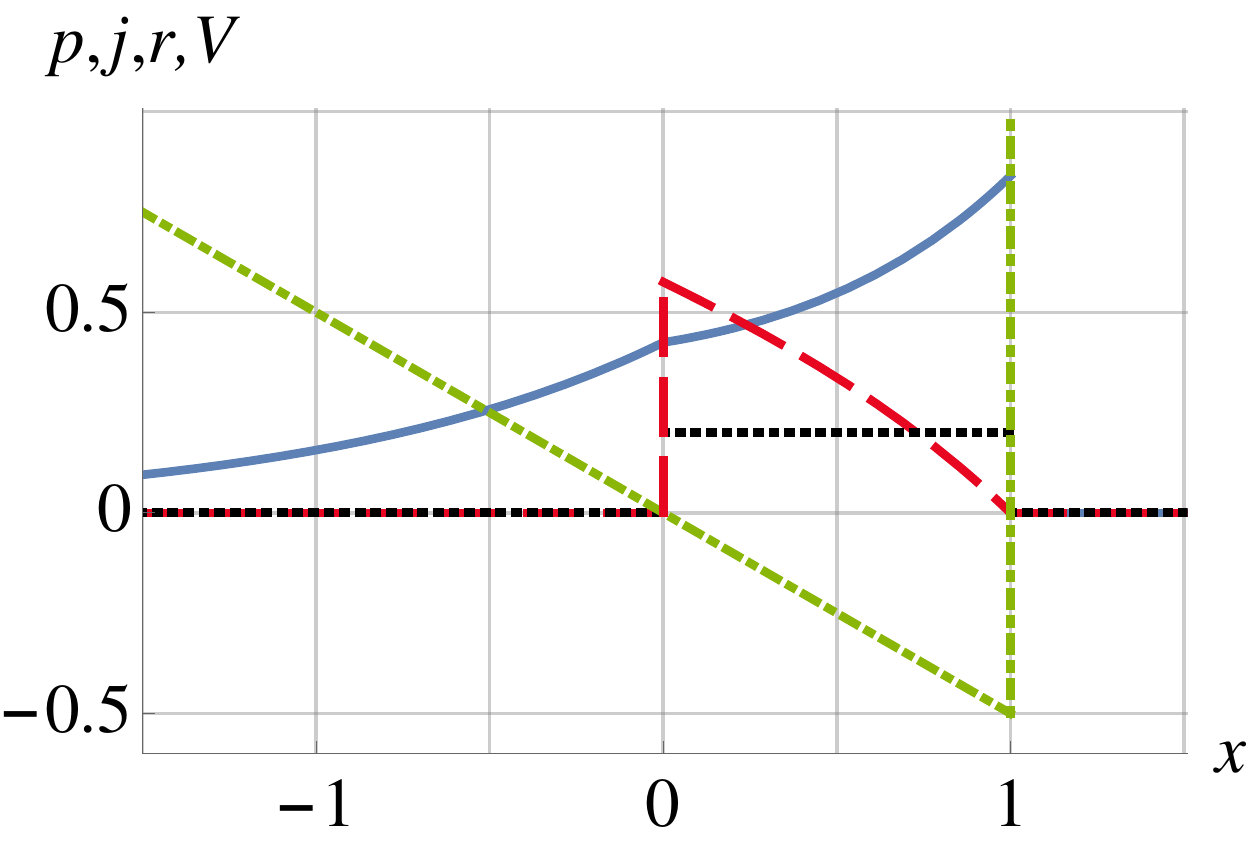}
      \label{fig:example_well}
    }
  \label{fig:examples}
  \caption{Three examples for reset mechanisms. We show the
    steady-state probability densities $p(x)$ (blue) and probability currents
    $j(x)$ (red) for varying potentials $V(x)$ (green) and resetting to
    $x_0=0$ at rates $r(x)$ (black).}
\end{figure}

\subsection{Free diffusion with resetting}

First, we choose $r(x)=r$ for $|x|\ge a$ and $r=0$ otherwise. The ensuing steady
state has been solved by Evans and Majumdar~\cite{Evans2011} and is shown in
fig.~\ref{fig:example_window}. Since $F=0$, there is no thermodynamic entropy
production $\Smed=0$. The entropy rates related to resetting,
eqs.~\eqref{eq:Sabs_cont} and~\eqref{eq:Sins_cont}, are
all non-zero and given by
\begin{align}
  \Sabs&=\frac{-2r}{2+a^2r+2a\sqrt{r}}\left(1-\ln \frac{\sqrt{r}}{2+a^2r+2a\sqrt{r}}\right),\\
  \Sins&=\frac{-2r}{2+a^2r+2a\sqrt{r}}\ln \frac{(1+\sqrt{r}a)\sqrt{r}}{2+a^2r+2a\sqrt{r}}.
\end{align}
The key point here is that the resetting counteracts the free diffusion by
confining the particle to a region around the origin. This reduction in the
uncertainty of the particle position is reflected by the rate of resetting
entropy production~\eqref{eq:Srst_cont} being strictly smaller than zero,
$\Srst<0$.

\subsection{Diffusion in a V-shaped potential}

We now apply a force $F(x)=\sgn(x)f$ to the colloidal particle which corresponds
to a roof-top potential ($f>0$) or a V-shaped potential ($f<0$) as shown in
fig.~\ref{fig:example_v}. The particle is reset to $x_0=0$ at a constant rate $r(x)=r$,
such that the Fokker-Planck equation~\eqref{eq:fpe} simplifies to
\begin{equation}
  \label{e5}
  \partial_x^2p(x)-\sgn(x) f\partial_xp(x)-rp(x)+ r\delta(x)=0
\end{equation}
We first solve eq.~\eqref{e5} for $x\neq0$. Using the continuity of $p(x)$ at
$x=0$ and imposing natural boundary conditions leads to
\begin{equation}
  p(x)=c \exp\left(-\frac{f+\sqrt{f^2+4r}}{2}|x|\right)
\end{equation}
with a constant c. Integrating eq.~\eqref{e5} from $-\epsilon$ to $\epsilon$
yields $c=(f+\sqrt{f^2+4r})/4$. 
This distribution and the corresponding probability current $j(x)$ are plotted in
fig.~\ref{fig:example_v}. The entropic rates then follow as
\begin{align}
  \Sabs&=r\ln \frac{f+\sqrt{f^2+4r}}{4}-r,\\
  \Sins&=-r\ln \frac{f+\sqrt{f^2+4r}}{4},\\
  \Smed&=\frac{-2fr}{f+\sqrt{f^2+4r}}.
\end{align}
The resetting entropy production rate obeys $\Srst=-r<0$, so again resetting reduces
the uncertainty of the system. The thermodynamic entropy production depends on
the potential shape, determined by the sign of $f$. For the roof-top potential,
heat is dissipated into the medium ($\Smed>0$) and work needs to be performed on
the system, $\Wout<0$. For the V-shaped potential, heat is absorbed from the surrounding
medium, $\Smed<0$, while work is extracted. Thus in the latter case, the
external reset mechanism operates like a Maxwell's demon, extracting heat from a
bath at constant temperature. From the perspective of the second law,
however, this is compensated by the work $\dWext$ required to maintain the resetting in
the steady state \eqref{eq:landauer}.
Hence the resetter has to apply more work than can be extracted as $\Wout$ from the
particle, ensuring thermodynamic consistency.

\subsection{Resetting with $\Srst>0$}

\begin{figure}
  \centering
  \includegraphics[width=.4\textwidth]{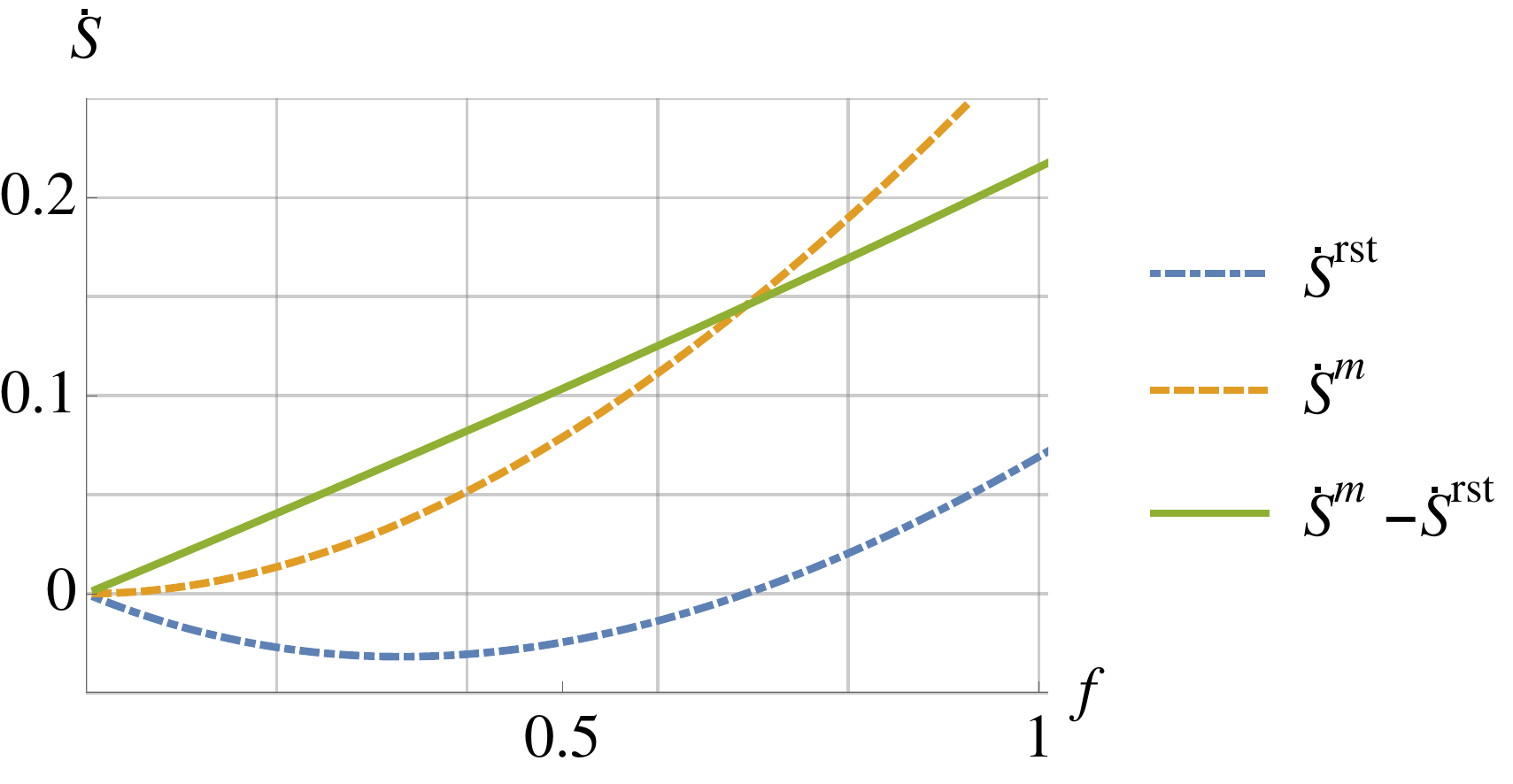}
  \caption{\label{fig:example_well_detailed} Resetting entropy production
    rate $\Srst$ (blue, dashed line) and thermodynamic entropy production
    rate $\Smed$ (yellow, dashed line) as a function of $f$ for the setup shown in
    fig.~\ref{fig:example_well} with $r=1$.}
\end{figure}

So far, the resetting entropy production has been negative in all the
examples. However, inspection of the rate, eq.~\eqref{eq:Srst_cont}, reveals
that it can also be positive if $p(x)>p(x_0)$ in regions of large $r(x)$, which can be realized in
a setup like the one shown in fig.~\ref{fig:example_well}. Here, we have a
reflecting boundary at $x=a$, hence $j(a)=0$, and apply a potential
$V(x)=-fx$ for $x \le a$. We reset the particle to $x_0=0$ at a rate $r(x)=r$
for $x \ge 0$ and $r(x)=0$ otherwise. The steady state distribution and current,
calculated analogously to the second example, are also shown in
fig.~\ref{fig:example_well}. The entropy production rates cannot
be calculated in closed form for such a system, but numerical results are shown
as a function of the force parameter $f$ in
fig.~\ref{fig:example_well_detailed}.  For $f$ larger than a critical value
$f_c$, the resetting entropy production rate $\Srst$ becomes indeed
negative, which corresponds to increasing the information content of the system.

\subsection{Operation diagram for the discrete random walk}

\begin{figure}
  \centering
  \includegraphics[width=0.3\textwidth]{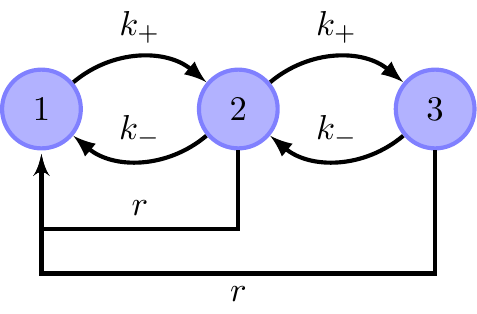}
  \caption{\label{fig:sketch_discrete} Three-state system with
      resetting from states $n=2,3$ to state $n_0=1$ with a state-independent
    reset rate~$r$.}
\end{figure}

\begin{figure}
  \centering
  \includegraphics[width=0.45\textwidth]{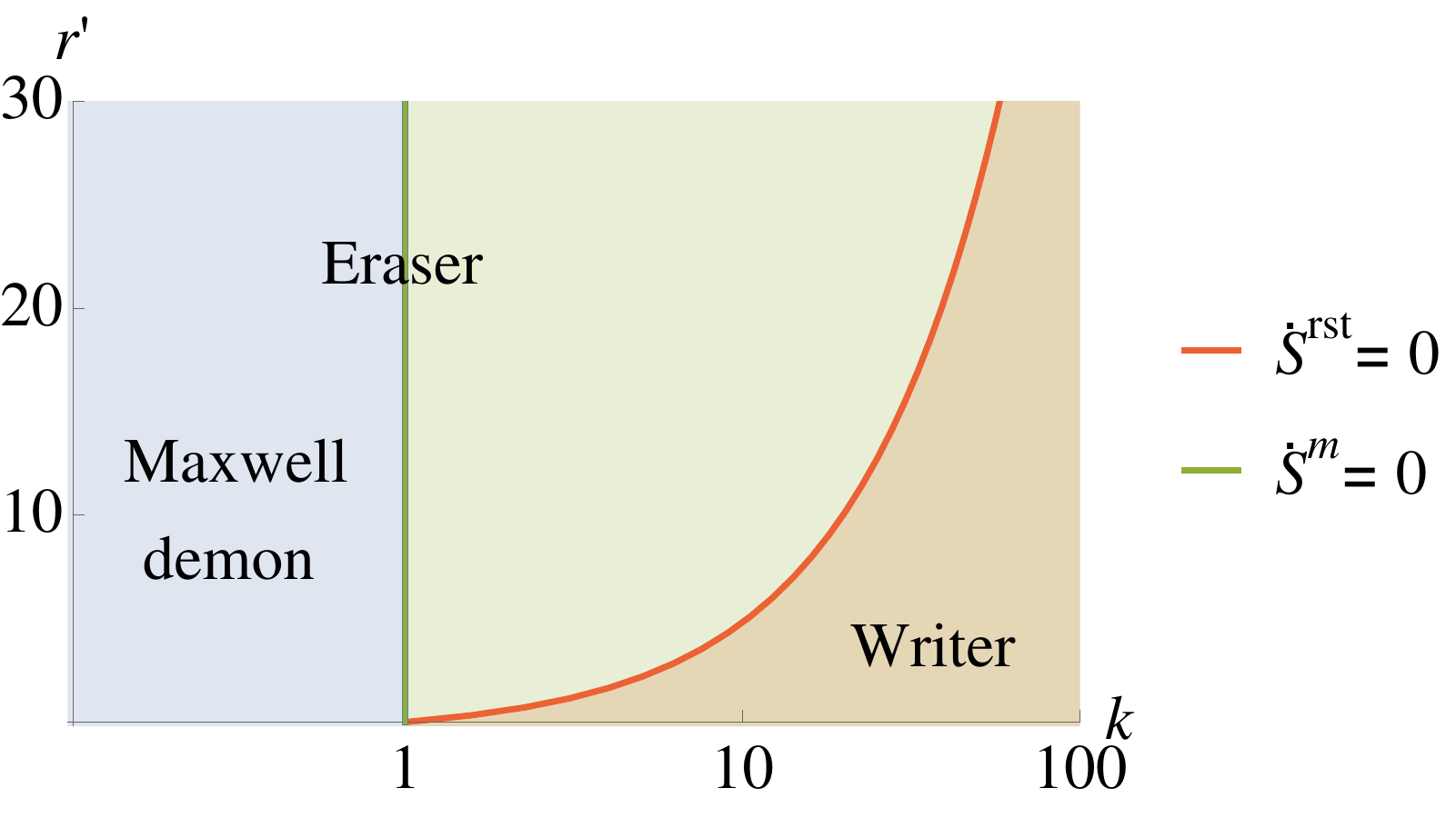}
  \caption{\label{fig:phase-diagram} Operation diagram for the system sketched in
    fig.~\ref{fig:sketch_discrete}. Three regimes emerge in the steady state which are plotted
     as a function of $k=k_+/k_-<1$ and the normalised reset rate
    $r'/k_-$. Heat is either extracted from ($\Smed<0$, blue domain), implementing
     a Maxwell demon, or dissipated into  the reservoir ($\Smed>0$).
     The resetting entropy production rate $\Srst$ is negative in
    the blue and green domain corresponding to continuously erasing information,
     but becomes positive in the brown domain corresponding to writing information.}
\end{figure}

In our last example, we expand on the idea of different signs for the
thermodynamic and resetting entropy production rates by considering the
three-state system shown in fig.~\ref{fig:sketch_discrete}. The transition rates
between neighbouring states are $k_+$ from left to right and $k_-$ in the
opposite direction. The random walker is reset from states $2$ and $3$ to the
initial state $n_0=1$ with constant rate~$r$. The master
equation~\eqref{eq:master} is then easily solved for the steady state, yielding
\begin{equation}
\begin{split}
p_1=&[k_+r+(k_- + r)^2]/N\\
p_2=&(k_-+r)k_+/N\\
p_3=&k_+^2/N
\end{split}
\end{equation}
with $N\equiv k_+(k_++r)+(k_-+r)(k_++k_-+r)$. In this system, three regimes
emerge with regard to the sign of the rates of entropy production in the medium
\eqref{eq:Smed_disc} and due to resetting~\eqref{eq:Srst_disc}. We sketch them
in the operation diagram, fig.~\ref{fig:phase-diagram}, as a function of the
parameters $k=k_+/k_-$ and $r'=r/k_-$. For $k<1$, as the random walker jumps to
the right, it absorbs heat $-\ln k>0$ from the reservoir which is extracted as
work $\Wout$ when it is reset to $n_0=1$, yielding an average thermodynamic
entropy production of $\Smed<0$, thus implementing a Maxwells demon. On the
other hand, for $k>1$, the random walker produces entropy in the medium at a
rate $\Smed>0$ and work has to be put in to reset the particle, yielding $\Wout<0$, effectively
lifting it to a higher state. Meanwhile, the resetting entropy rate
depends on the ratio of probabilities in the initial state $n_0=1$ and in states
$n=2,3$, respectively, either erasing or increasing the information content of the system.
However, for all $r'$ and $k$, $\Smed - \Srst > 0$.

\section{Conclusion and perspectives}

We have derived the first and second law of thermodynamics for continuous and
discrete stochastic dynamics with resetting and identified the contribution of
resetting to the total entropy production. This resetting entropy rate $\Srst$
quantifies the rate of information creation or erasure in the steady
state. Using Landauer's principle, we were able to derive a lower bound on the
work input required to implement a given resetting mechanism. We note that work
can also be extracted from the heat bath at constant temperature in a Maxwell's
demon type setup at the expense of increased external work input.

We expect that our results can be verified experimentally, for example using
colloids in an optical trap, which have a history of successful validation of
concepts from stochastic thermodynamics
\cite{Blickle2006,toyabe2010,Blickle2011,Berut2012}. Among the avenues for further
theoretical work, it should be worthwhile to apply our framework to
biological systems featuring resetting, such as the detection of pathogens by
the immune system \cite{Mora2015}. An appreciation of the thermodynamic costs
involved could guide the search for the fundamental principles underlying these
highly efficient processes.

\acknowledgements 

We thank D. Hartich and P. Pietzonka for careful reading of the manuscript and
D. Schmidt for support producing the figures.

\bibliographystyle{eplbib}
\bibliography{library}

\begin{thebibliography}{10}
\expandafter\ifx\csname url\endcsname\relax\def\url#1{\texttt{#1}}\fi

\bibitem{Evans2011a}
\Name{Evans M.~R. \and Majumdar S.~N.} \REVIEW{Phys. Rev.
  Lett.}{106}{2011}{160601}.

\bibitem{Kusmierz2014}
\Name{Kusmierz L., Majumdar S.~N., Sabhapandit S. \and Schehr G.} \REVIEW{Phys.
  Rev. Lett.}{113}{2014}{220602}.

\bibitem{Gupta2014}
\Name{Gupta S., Majumdar S.~N. \and Schehr G.} \REVIEW{Phys. Rev.
  Lett.}{112}{2014}{220601}.

\bibitem{meylahn2015}
\Name{Meylahn J.~M., Sabhapandit S. \and Touchette H.} \REVIEW{Phys. Rev.
  E}{92}{2015}{062148}.

\bibitem{eule2015a}
\Name{Eule S. \and Metzger J.} \REVIEW{preprint}{}{2015}{arxiv:1510.07876}.

\bibitem{Hopfield1974}
\Name{Hopfield J.~J.} \REVIEW{Proc. Natl. Acad. Sci.}{71}{1974}{4135}.

\bibitem{Ninio1975}
\Name{Ninio J.} \REVIEW{Biochimie}{57}{1975}{587}.

\bibitem{murugan2014}
\Name{Murugan A., Huse D.~A. \and Leibler S.} \REVIEW{Phys. Rev.
  X}{4}{2014}{021016}.

\bibitem{hartich2015a}
\Name{Hartich D., Barato A.~C. \and Seifert U.} \REVIEW{New J.
  Phys.}{17}{2015}{055026}.

\bibitem{Kyriakidis1994}
\Name{Kyriakidis E.} \REVIEW{Stat. Probab. Lett.}{20}{1994}{239}.

\bibitem{Dharmaraja2015}
\Name{Dharmaraja S., {Di Crescenzo} A., Giorno V. \and Nobile A.~G.} \REVIEW{J.
  Stat. Phys.}{161}{2015}{326}.

\bibitem{Landauer1961}
\Name{Landauer R.} \REVIEW{IBM J. Res. Dev.}{5}{1961}{183}.

\bibitem{Bennett1982}
\Name{Bennett C.} \REVIEW{Int. J. Theor. Phys.}{21}{1982}{905}.

\bibitem{Seifert2012}
\Name{Seifert U.} \REVIEW{Rep. Prog. Phys.}{75}{2012}{126001}.

\bibitem{Parrondo2015}
\Name{Parrondo J. M.~R., Horowitz J.~M. \and Sagawa T.} \REVIEW{Nat.
  Phys.}{11}{2015}{131}.

\bibitem{Berut2012}
\Name{B{\'{e}}rut A., Arakelyan A., Petrosyan A., Ciliberto S., Dillenschneider
  R. \and Lutz E.} \REVIEW{Nature}{483}{2012}{187}.

\bibitem{Jun2014}
\Name{Jun Y., Gavrilov M. \and Bechhoefer J.} \REVIEW{Phys. Rev.
  Lett.}{113}{2014}{190601}.

\bibitem{lan2012}
\Name{Lan G., Sartori P., Neumann S., Sourjik V. \and Tu Y.} \REVIEW{Nat.
  Phys.}{8}{2012}{422}.

\bibitem{barato2014a}
\Name{Barato A.~C., Hartich D. \and Seifert U.} \REVIEW{New J.
  Phys.}{16}{2014}{103024}.

\bibitem{hartich2016}
\Name{Hartich D., Barato A.~C. \and Seifert U.} \REVIEW{Phys. Rev.
  E}{93}{2016}{022116}.

\bibitem{Evans2011}
\Name{Evans M.~R. \and Majumdar S.~N.} \REVIEW{J. Phys. A Math.
  Theor.}{44}{2011}{435001}.

\bibitem{Pal2015}
\Name{Pal A.} \REVIEW{Phys. Rev. E}{91}{2015}{012113}.

\bibitem{Gardiner1983}
\Name{Gardiner C.} \Book{{Stochastic Methods}} (Springer, Berlin-Heidelberg-New
  York-Tokyo) 1983.

\bibitem{seifert2005a}
\Name{Seifert U.} \REVIEW{Phys. Rev. Lett.}{95}{2005}{040602}.

\bibitem{Esposito2011}
\Name{Esposito M. \and {Van den Broeck} C.} \REVIEW{EPL (Europhysics
  Lett.}{95}{2011}{40004}.

\bibitem{mandal2012}
\Name{Mandal D. \and Jarzynski C.} \REVIEW{Proc. Natl. Acad. Sci. U. S.
  A.}{109}{2012}{11641}.

\bibitem{barato2013b}
\Name{Barato A.~C. \and Seifert U.} \REVIEW{EPL (Europhysics
  Lett.}{101}{2013}{60001}.

\bibitem{barato2014}
\Name{Barato A.~C. \and Seifert U.} \REVIEW{Phys. Rev.
  Lett.}{112}{2014}{090601}.

\bibitem{Blickle2006}
\Name{Blickle V., Speck T., Helden L., Seifert U. \and Bechinger C.}
  \REVIEW{Phys. Rev. Lett.}{96}{2006}{070603}.

\bibitem{toyabe2010}
\Name{Toyabe S., Sagawa T., Ueda M., Muneyuki E. \and Sano M.} \REVIEW{Nat.
  Phys.}{6}{2010}{988}.

\bibitem{Blickle2011}
\Name{Blickle V. \and Bechinger C.} \REVIEW{Nat. Phys.}{8}{2011}{143}.

\bibitem{Mora2015}
\Name{Mora T.} \REVIEW{Phys. Rev. Lett.}{115}{2015}{038102}.

\end{thebibliography}

\end{document}